# Layer breathing modes in few-layer graphene

Chun Hung Lui and Tony F. Heinz*

*Departments of Physics and Electrical Engineering, Columbia University, 538 West 120th Street, New York, NY 10027, USA*

*e-mail: tony.heinz@columbia.edu

Layer-layer coupling plays a critical role in defining the physical properties of few-layer graphene (FLG). With respect to vibrations, the interlayer coupling is predicted to create a set of $N$-1 finite frequency, out-of-plane interlayer modes in $N$-layer graphene. Unlike the widely studied in-plane vibrations, the properties of these layer-breathing modes (LBMs) are defined by the layer-layer interactions. Here we report direct observation of the distinct LBMs for graphene of layer thicknesses $N = 2 – 20$ through electronically resonant overtone Raman bands observed in the range of 80-300 cm$^{-1}$. The Raman bands exhibit multi-peak features that are unique for graphene samples of each layer thickness up to 20 layers. The frequencies of the set of layer breathing modes for all samples can be described within a simple model of nearest plane coupling.

Mono- and few-layer graphene (FLG) are highly attractive materials because of their distinctive physical properties and potential for novel applications (1-3). A striking feature of FLG materials is that each layer thickness possesses a unique low-energy electronic structure, with interlayer coupling yielding a rich spectrum of electronic properties for the materials (1-9). Similar distinct features are also predicted with regard to phonons, and a set of low-energy *interlayer* vibrational modes are predicted with distinct layer thickness dependence. These interlayer vibrations are comprised of shearing modes (10) and layer breathing modes (11-17) (LBMs, or ZO', Fig. 1A). These vibrations involve, respectively, the relative displacement of the individual graphene layers in the tangential and vertical direction. The LBMs are of great scientific and practical significance because the out-of-plane layer displacements make them highly sensitive to interaction between the graphene layers. Although the LBM was studied in bulk graphite (18, 19) and also explored theoretically in FLG (11-17), direct observation of the LBM vibrations in FLG has proven to be challenging because of the low frequency of the vibrations and the atomic-level thickness of the samples. To date, experimental signatures of LBMs have been obtained only through Raman spectroscopy of LO+ZO' combination modes (20-22). Detailed study of the low-frequency LBMs through this approach has, however, been impeded by the much larger vibrational frequency of the in-plane longitudinal optical (LO) phonon involved in the combination mode, which renders precise extraction of the frequencies of the LBM difficult.

Here we report a determination of the different LBM frequencies for FLG by means of the two-phonon overtone spectra observed in doubly resonant Raman spectroscopy. We find for graphene samples of layer thickness from 2 to 20 layers, over the spectral range of 80-300 cm$^{-1}$, features arising from *each of the different LBM vibrations*. These numerous 2ZO' Raman features shift systematically as the layer thickness of graphene increases. The behavior reflects the increasing number and the shifting frequencies of the different ZO' normal modes with increasing graphene layer thickness. We show that these frequencies can be described surprisingly accurately using a simple model based on nearest-neighbor couplings between the layers, in essence a one-dimensional chain of $N$ coupled masses.

The use of the 2D-mode Raman line shape to determine the thickness and stacking order of graphene layers has, we should note, become a standard component of graphene research (23, 24). Since this response arises from phonons that are selected by an electronic resonance (25, 26), the Raman spectra read out information about the electronic transitions through the phonon dispersion. The method does not rely on an intrinsic difference in the dispersion relation of the in-plane phonons with layer thickness. As a consequence, the differences in the spectra become very subtle for more than 3 or 4 layers of graphene. The differences in the Raman spectra of the LBM presented in this paper, however, arise from a completely distinct mechanism, namely, the *strong variation in the character and frequency of the underlying phonons* as a function of layer thickness. The intrinsic properties of the LBM phonons, like those of the recently reported shearing mode phonons (10), change with each additional layer of graphene. Measurement of these phonons thus allows direct determination of layer thickness, with single-layer precision, up to 20 layers. We also expect that these out-of-plane modes will exhibit enhanced sensitivity to coupling to the external environment.

We performed Raman measurements on high-quality free-standing graphene layers of thickness from single layer to 20 layers (see Methods). Let us first consider the Raman response of bilayer graphene (2LG). This material should exhibit only a single LBM, corresponding to the symmetric oscillation of the two sheets of graphene about the center of the structure (Fig. 1A). Over a frequency range of 145 – 220 cm$^{-1}$, we observed a double-peak feature at ~180 cm$^{-1}$ (Fig. 1B). Although the intensity of this feature is approximately 100 times weaker than that of the G-mode response of the in-plane vibrations, these low-frequency peaks are clearly observable. On the other hand, when we examine the Raman spectrum of single-layer graphene, we see no features in this spectral region. This result implies that the observed mode corresponds to an interlayer vibration. As the laser photon energy $E_{exc}$ increases from 1.58 eV to 2.33 eV, the Raman band of 2LG significantly blue-shifts. The line shape also changes, with the two components separating further from one another.

The strongly dispersive Raman response indicates that we are observing a higher-order Raman process, like the 2D-mode, which is known to be highly efficient for graphitic materials through mechanisms involving electronic resonances (23-26). We have analyzed the dispersion behavior of the measured 2LG Raman response using double-resonance theory. We assign the features to the overtone (2ZO') mode of the layer breathing vibration (Fig. 1A) occurring through an intra-valley resonance processes. In our treatment, we adopt the electronic band structure of 2LG within a simplified tight-binding (TB) model that includes just the dominant intra- ($\gamma_0$) and inter-layer ($\gamma_1$) couplings. (We note, however, that analysis using more accurate band structure does not significantly modify the results.) 2LG has two conduction (valence) bands, $c1$ and $c2$ ($v1$ and $v2$), which give rise to four possible electronically resonant processes (P11, P22, P12 and P21) for the intra-valley two-phonon scattering (Fig. 1C). Group analysis (27) shows that P12 and P21 are forbidden at the high symmetry points along the $\Gamma$-$M$ and $M$-$K$ lines in the Brillouin zone (BZ) of 2LG, thus suppressing these processes. We can then assign the high- and low-energy component peaks (2ZO'$^+$ and 2ZO'$^-$) of the observed Raman feature, respectively, to the P11 and P22 processes. With the aid of these assignments we can estimate the dominant phonon momenta in the resonant Raman processes for different values of $E_{exc}$. Fig. 2D displays the phonon energies, obtained as half of the values of the 2ZO'$^+$ and 2ZO'$^-$ peaks, as a function of the phonon momentum. We compare the experimental results with the predicted dispersion relation for the ZO'-mode in 2LG (16). The good agreement between experiment and theory confirms our assignment of the 2ZO' overtone mode. From the results we can extrapolate the ZO'-mode frequency for 2LG at the $\Gamma$-point to be 80 ± 2 cm$^{-1}$, a result comparable to theoretical predictions (12, 16). This latter frequency corresponds to that of oscillation in the relative displacement of two *rigid* graphene sheets in the bilayer. We note that the only other interlayer mode present for 2LG is the shearing mode. Its frequency, however, for the range of wave vectors relevant in the double-resonance process, is expected to lie at considerably higher frequency (TA and LA branches in Fig. 1D).

We now extend our discussion to the behavior for thicker samples of FLG. As additional layers of graphene are added beyond the bilayer, new normal modes develop. For *N*-layer graphene (NLG), the interlayer coupling will create *N*-1 distinct ZO' branches (with finite frequencies at the zone center). We denote these modes as ZO'$_N^{(n)}$ (or ZO'$^{(n)}$ when $N$ is clear from the context), where the index $n$ = 1, 2, …, *N*-1 represents the branch number, enumerating from high to low mode frequency. The highest frequency mode will correspond to alternating displacements of adjacent graphene planes, while lower frequency modes will involve more layers moving in the same direction and experiencing consequently a weaker overall restoring force. Remarkably, as we show below, we are able to *observe all of these different LBMs* within our experimental spectral range for samples up to 20 layers in thickness.

As an illustration of the evolution of behavior with increasing layer thickness, we consider 4LG. 4LG possesses three relevant ZO' branches: ZO'$_4^{(1)}$, ZO'$_4^{(2)}$ and ZO'$_4^{(3)}$, whose layer displacements are shown schematically in Fig. 2A. In accordance, we observe three Raman features in the spectral range of 85-270 cm$^{-1}$ for $E_{exc}$ = 2.33 eV (Fig. 2B). Each features shows a double peaks, with a position and line shape evolving in a manner similar to that of the 2ZO' mode in 2LG as $E_{exc}$ is varied. To confirm the origins of these Raman bands, we adopt the same analysis as in 2LG, but using 4LG TB electronic structure with couplings $\gamma_0$ and $\gamma_1$. Since the 4LG electronic structure exhibits four conduction (valence) bands, the electronic processes in the intravalley two-phonon resonant Raman scatterings are numerous and complex. For simplicity, we assume that the high- and low-frequency components of the double-peaked Raman features arise from electronic processes with the maximum and minimum phonon momentum, respectively. We then obtain the phonon frequencies as a function of momentum (Fig. 2C). We find that the experimental results match well with the theoretical dispersion (16) of the ZO'$_4^{(1)}$, ZO'$_4^{(2)}$ and ZO'$_4^{(3)}$ phonon modes. Similar results hold for analysis on 3LG.

For graphene samples of still greater thickness, we observe the emergence of new Raman features, corresponding to the presence of additional LBMs. For 11LG (Fig. 3A-C), for example, we have 10 LBMs, ZO'$_{11}^{(1)}$ - ZO'$_{11}^{(10)}$. We observe 8 distinct Raman peaks in the spectral range of 85 – 300 cm$^{-1}$ for $E_{exc}$ = 2.33 eV. As we show below by quantitative analysis, we can identify these features with overtones of the ZO'$_{11}^{(1)}$ - ZO'$_{11}^{(8)}$ phonon branches (Fig. 3C). The two lowest-frequency phonon modes, ZO'$_{11}^{(9)}$ and ZO'$_{11}^{(10)}$, are not present because of the restricted spectral range of observation.

We have examined the Raman spectra of suspended graphene samples with each thickness $N$ = 2, 3, 4 … 20, as well as bulk graphite, all with Bernal stacking. In the frequency range of 80 – 300 cm$^{-1}$, we find panoply of Raman peaks. The observed frequencies and line shapes are unique for graphene of each layer thickness (Fig. 4A). With increasing layer thickness, the number of 2ZO' Raman peaks grows systematically and the spectral shape of the whole band gradually approaches that of the graphite spectrum. The above observations highlight the remarkable sensitivity of the LBMs to interlayer interactions. In particular, the 2ZO' peaks are still distinct for 20LG, indicating that they can be used as an accurate spectroscopic signature to characterize, with atomic level precision, the layer thickness of graphene up to 20 layers.

We note that multiple sub-peaks (denoted by dots of the same color in Fig. 4A-B) are observed for some 2ZO' Raman features. They are associated with the same ZO' phonon branch, but with phonons of different momenta, in agreement with our earlier analysis of the double-resonance effects on the 2LG and 4LG spectra. The high-frequency sub-peaks gradually subside as $N$ increases. We thus generally observe only the low-frequency sub-peak for each LBM in FLG with $N$ > 8. (see Supplementary Information for detailed discussions and a comparative study of 2ZO' mode for 3LG with ABA and ABC stacking to show the role of electronic structure.)

We now show that the frequencies of the different LBM vibrations can be described well by a remarkably simple model, namely, one in which all layers are treated as equivalent and only nearest-neighbor interactions between the layers are considered. Within this picture, the various LBM modes ZO'$_N^{(n)}$ are equivalent to a linear chain of $N$ masses connected by springs. The solution to this classic problem yields mode frequencies of (28)

$$\omega_N(n) = \omega_o \sin[(N-n)\pi/2N], \quad [1]$$

where $\omega_o$ denotes the frequency of the bulk optical mode or, equivalently, $\omega_o/\sqrt{2}$ corresponds to the frequency of the LBM in the bilayer. To adapt this simple analysis to our problem, we chose $\omega_o$ = 132.3 cm$^{-1}$ as the ZO' mode frequency observed in bulk graphite, as obtained from the measured frequency of the 2ZO' Raman peak (264.5 cm$^{-1}$) (Fig. 4A). As the 2ZO' mode involves phonons of finite momentum through the double-resonance Raman process, a minor error (~2 cm$^{-1}$) is expected from the in-plane phonon dispersion of the LBMs near the zone center.

The results of Eq. (1) provide an excellent overall fit to the experimental data for all layer thicknesses (Fig. 4B). The results for the case of 11LG (Fig. 3C) highlight the good agreement that is achieved. The modest observed deviations presumably reflect the approximate treatment of the finite in-plane wave vector in the data, as well as limitations in the model, such as considering the interactions between all graphene layers to be completely identical, irrespective of layer thickness or position. Since suspended graphene samples were investigated, we expect that perturbations of the external environment will be negligible.

For further analysis on the LBMs in FLG, we note that, within a model of nearest-layer interactions, the frequencies of the various ZO'$_N^{(n)}$ modes are the same as frequencies of the bulk graphite ZO' mode for out-of-plane wave vectors of $(\pi/c)(N-n)/N$ along the $\Gamma$-$A$ line in the three-dimensional graphite BZ, where $c \sim 0.34$ nm is the interlayer spacing. We have applied this notion to reconstruct the expected graphite phonon dispersion along the $\Gamma$-$A$ line using the observed 2ZO' modes in FLG (Fig. 4C). We find good overall agreement between experiment and existing calculations (16). The experimental results appear to be slightly higher than the theoretical values as expected because of the finite in-plane phonon dispersion of the LBMs selected in the doubly-resonant Raman measurements.

Finally we comment on possibility of combination modes of ZO' phonons from different branches in our observed Raman spectra. We may perceive their role from the graphite spectrum (Fig. 4A), which exhibits two prominent peaks at 180 and 265 cm$^{-1}$. The latter can be readily interpreted as the 2ZO' overtone band arising from a singularity of the LBM phonon density of state (DOS) at the $\Gamma$ point (Fig. 4C). The former cannot, however, be explained in this way due to the lack of similar singularity in this energy range. This leads us to consider two-phonon combination modes. Since the two scattered phonons must have equal (but opposite) momentum in the out-of-plane ($\Gamma$-$A$) direction, the only allowed combinations are of one LO- and one LA-branch LBM phonon with the same momentum (Fig. 4C). The resultant LO+LA combination modes will be of frequencies close to double that of the $A$-point LBM phonon frequency, where a singularity of the two-phonon DOS occurs. This accounts naturally for the Raman peak at 180 cm$^{-1}$ in graphite. The above selection rules should still be approximately valid for FLG. Correspondingly, we would expect significant Raman response only from ZO'$_N^{(n)}$ + ZO'$_N^{(N-n)}$ combination modes, which would satisfy the condition of vanishing total phonon momentum in the bulk limit. Such combination modes are limited to a spectral range around 150 - 200 cm$^{-1}$. In this region, we do see somewhat enhanced Raman peaks, presumably arising from a contribution of these combination modes in addition to the dominant overtone response (Fig. 4A). Such propensity rules for the combination modes are supported by the complete absence of any peaks in the experimental Raman spectra at frequencies not compatible an overtone mode, such as would occur for a ZO'$_N^{(n)}$ + ZO'$_N^{(n+1)}$ combination mode.

In conclusion, we have investigated, through two-phonon Raman scattering, the layer-breathing vibrations in few-layer graphene. We observe, for each layer thickness, a whole family of layer breathing normal modes, with a completely distinct vibrational spectrum from one another. This seemingly complex and layer-dependent behavior of LBMs can be described well within a model based on nearest-neighbor couplings between layers. In addition, our research provides a powerful optical means of characterizing the layer number of graphene, with atomic-level precision, over an unprecedented range of thickness up to 20 layers. More generally, the LBMs, with their out-of-plane layer displacements, provide a route for probing the influence of adsorbates, surrounding media, and other environmental perturbations on few-layer graphene. The distinctive characteristics of LBMs demonstrated in our work should also extend analysis of other two-dimensional layered materials, such as the hexagonal boron nitride and transition metal dichalcogenides.

## Methods

**Sample preparation and characterization** In our experiment, we investigated pristine free-standing graphene layers prepared by the mechanical exfoliation of kish graphite (Covalent Materials Corp.). The free-standing samples were deposited on quartz substrates with pre-patterned trenches (4 μm width x 2 μm depth). Using a confocal arrangement to collect the Raman signal, we were able to discriminate against inelastic light scattering from the substrate. In addition, the suspended samples allowed us to investigate the LBM without the influence of external perturbations from interactions of the FLG samples with the substrate. We examined graphene with thicknesses from single layers up to 20 layers, all with the Bernal stacking order. We also studied suspended 3LG with both ABA (Bernal) and ABC (rhombohedral) stacking order (see Supplementary Information). The layer thickness and stacking order were characterized by several methods, including infrared spectroscopy (9, 29), optical contrast measurements in the visible range, Raman spectroscopy of the 2D-mode (23, 24, 30) and the LOZO' combination mode (20, 21), and atomic force microscopy (AFM).

**Raman spectroscopy of LBMs** The Raman measurements were performed using a commercial confocal microRaman system (JY-Horiba labRAM ARAMIS) equipped with a cooled charge-coupled device (CCD) array detector. Laser excitation sources were available with photon energies (wavelengths) of $E_{exc}$ = 1.58 eV (785 nm), 1.96 eV (633 nm), and 2.33 eV (532 nm). Since the Raman response of the 2ZO' mode is relatively weak and involves a small frequency shift, we performed the measurements under an argon atmosphere to suppress the rotational Raman lines of the air. In order to record weak Raman features, we collected data for periods of over an hour. The laser power was consistently maintained below 5 mW, with a spot size of a few microns on the sample. As heat dissipation is reduced for suspended graphene, we observed some heating effect in our samples. This was estimated to cause a red shift of ~2 cm$^{-1}$ for the 2ZO' Raman peak.

**Acknowledgements** We thank K. Knox, S. H. Kim and D. Q. Zhang for assistance in sample and substrate preparation, L. M. Malard for discussions, L. J. Karssemeijer and A. Fasolino for sharing their theoretical calculation of the low-energy phonon dispersion of FLG and graphite. We also acknowledge support from the Office of Naval Research under the MURI program, from DOE Basic Energy Sciences under grant FG02-98ER14861, from the National Science Foundation under grant CHE-0117752 and the NRI program of the SRC, and from the New York State Office of Science, Technology, and Academic Research (NYSTAR).

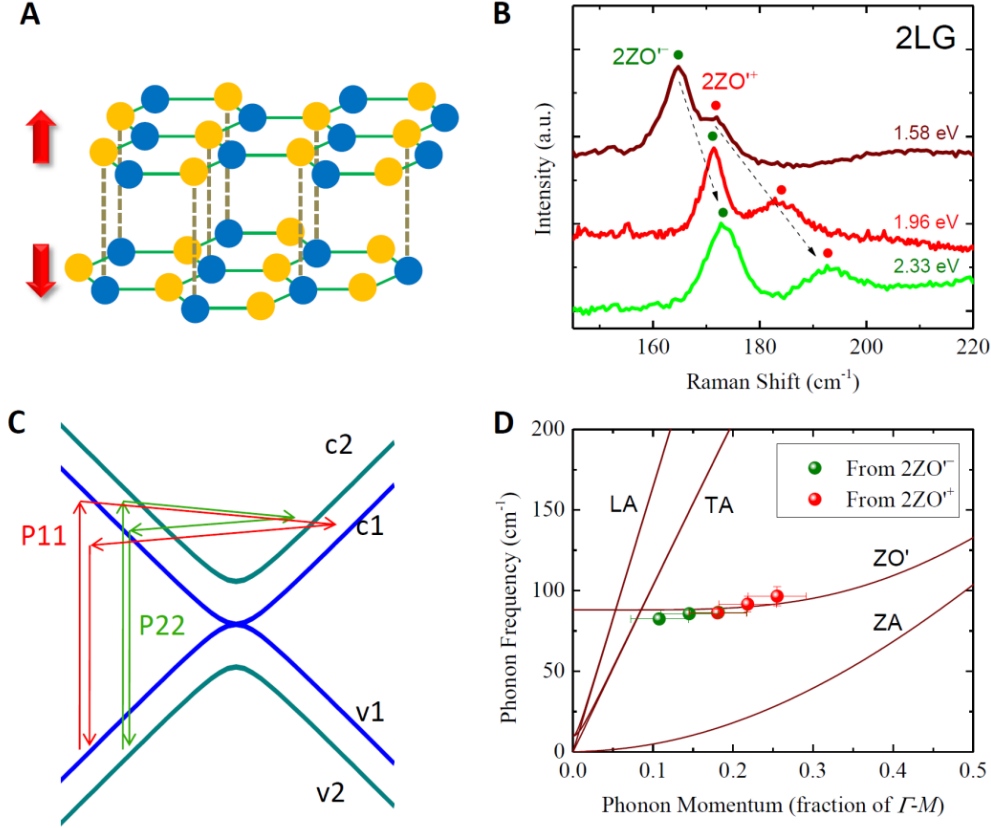

**Fig. 1.** The overtone layer-breathing mode (2ZO') for bilayer graphene. (A) Schematic atomic displacement of LBM of 2LG. (B) Normalized 2ZO'-mode spectra of suspended 2LG at different excitation laser energies $E_{exc}$. The high- and low-energy component of the 2ZO' band are denoted as 2ZO'$^{+}$ and 2ZO'$^{-}$, respectively. (C) The main electronic scattering processes in the two-phonon double-resonance Raman mechanism of the 2ZO' mode in 2LG. (D) The ZO'-mode phonon dispersion (symbols) extracted from (B). The phonon frequency is obtained as half of the value of the Raman shift for the peaks in (B) through the fits by a double Gaussian function. The error bars reflect the FWHMs of the fit Gaussian functions. The 2ZO'$^{+}$ and 2ZO'$^{-}$ peaks are ascribed to the P11 and P22 processes in (C), respectively. The corresponding error bars represent uncertainties in estimating the phonon momenta in these scattering processes. The experimental data are compared with the theoretically predicted dispersion (16) for the LBM in 2LG (lines).

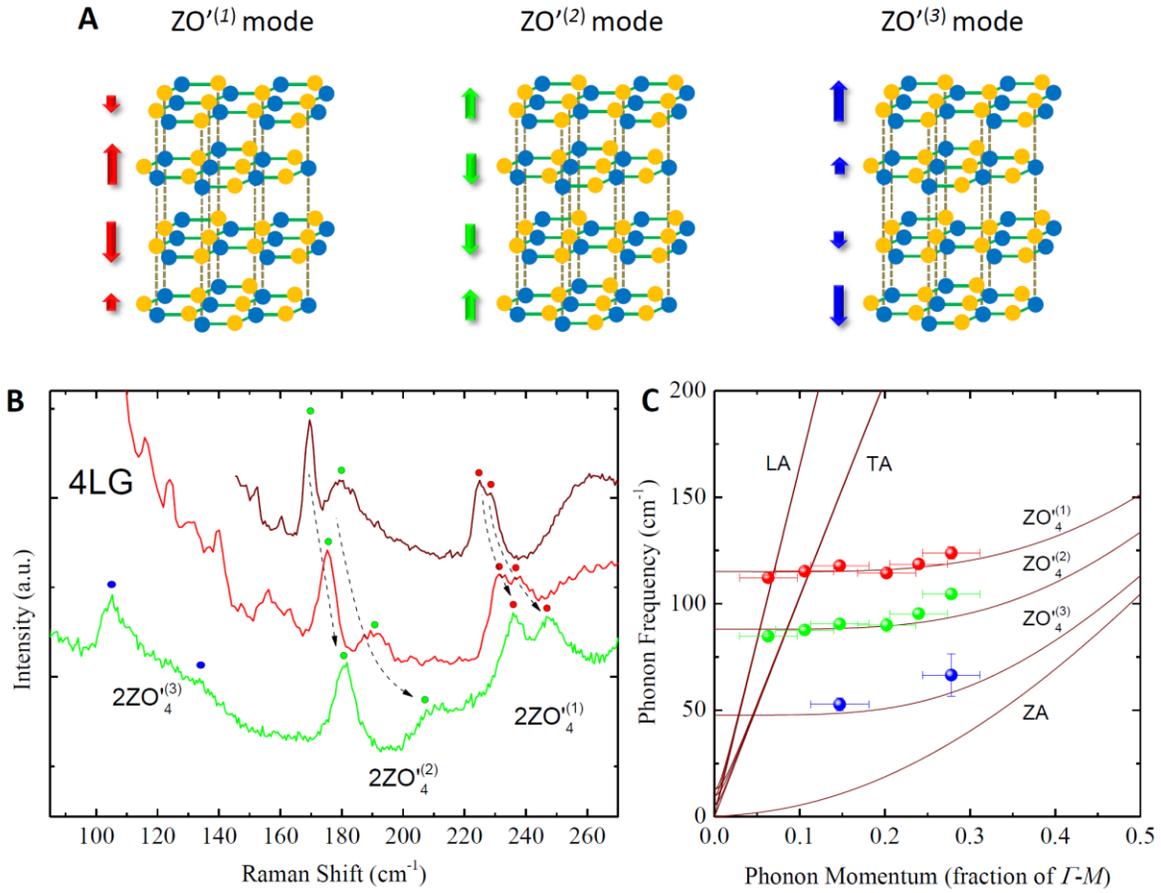

**Fig. 2**. The overtone layer breathing modes for 4LG. (A) Schematic atomic displacement of the three ZO' modes in 4LG. The length of the arrows represents the magnitude of the vibration. (B) Raman spectra of the three 2ZO' modes in suspended 4LG for $E_{exc}$ = 1.58 (upper trace), 1.96 (middle trace) and 2.33 eV (lower trace). (C) ZO'-mode phonon dispersion, extracted from (B) as in the case of 2LG. In determining the phonon momenta, we assume that the high (low)-frequency components of the overtone modes arise from electronic processes with maximum (minimum) phonon wave vector defined by the band structure of 4LG for the double-resonance Raman scattering. The experimental data are compared with the theoretically predicted dispersion (16) of the low-frequency phonon modes in 4LG.

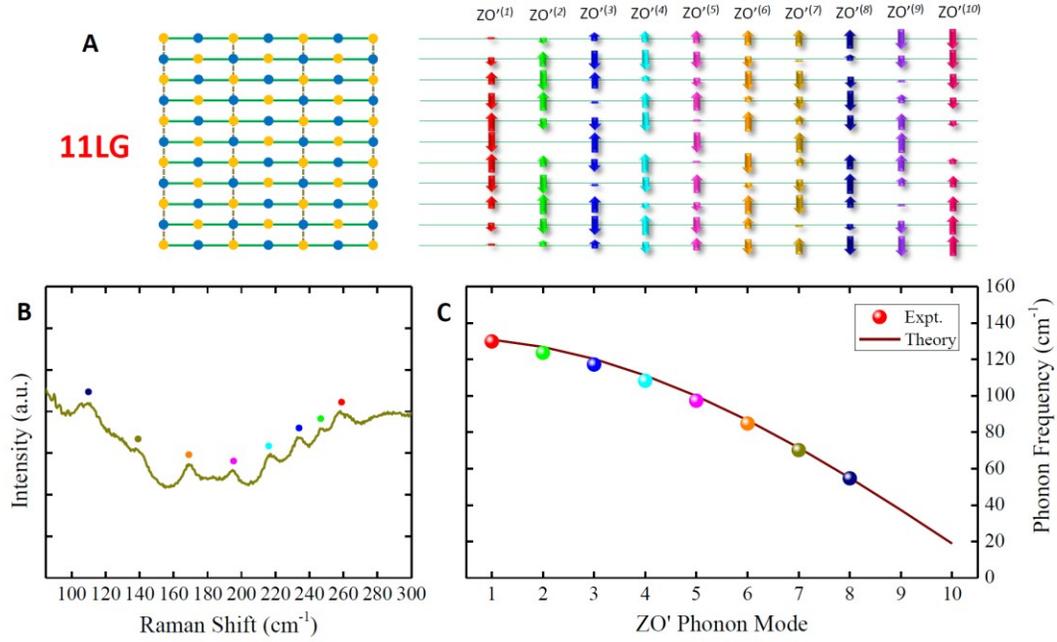

**Fig. 3.** The overtone layer breathing modes for 11LG. (A) Schematic atomic displacement of the ten ZO' modes in 11LG. The left panel shows the layer structure of 11LG sample. The right panel shows the layer displacement in each ZO' mode, with arrows drawn in proportion to the displacement magnitude. (B) The 2ZO'-mode Raman spectrum of suspended 11LG for $E_{exc}$ = 2.33 eV. (C) The ZO'-mode phonon frequency, obtained as half of the frequency of the corresponding Raman peaks in (B). The experimental data are compared with the predictions of the model described in the text (Eq. 1) for the ZO'-mode frequencies for 11LG.

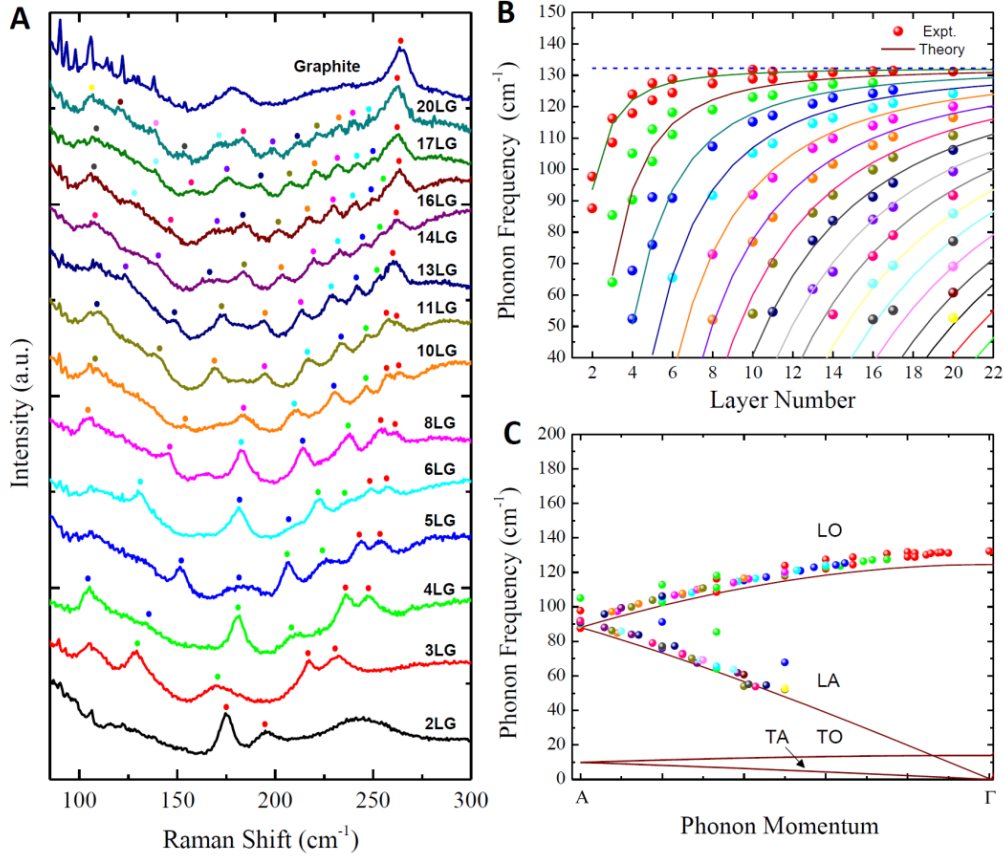

**Fig. 4.** The layer breathing modes for few layer graphene. (A) The 2ZO'-mode Raman spectra for suspended graphene samples of layer thickness from $N = 2$ to 20 and for the bulk graphite. The spectra are up-shifted successively for clarity. The sharp features below 150 cm$^{-1}$ in the graphite spectrum arise from the Raman scattering in air. (B) The ZO'-mode phonon frequencies, obtained from the corresponding Raman peaks in (A), as a function of layer thickness. The experimental results (symbols) are compared with theory (lines) as described in the text (Eq. 1). (C) The LBM phonon dispersion along the $A$-$\Gamma$ line in graphite, constructed from the data in (B) within a zone-folding scheme, in comparison with the theoretically predicted dispersion (16) of the low-frequency phonons in graphite.

# Supplementary Information of
# "Layer breathing modes in few-layer graphene"

Chun Hung Lui and Tony F. Heinz

*Departments of Physics and Electrical Engineering, Columbia University, 538 West 120th Street, New York, NY 10027, USA*

1. **The multiple sub-peaks in the 2ZO' Raman band**

According to Figure 4A-B in the main paper, multiple sub-peaks are observed in some 2ZO' Raman features of few-layer graphene (FLG). In accordance with our analysis of the 2LG and 4LG spectra in the main text based on the double-resonance Raman theory (1-5), these different sub-peaks, as denoted in the spectra by dots of the same color, correspond to ZO' phonons of different momenta within the same ZO'$_N^{(n)}$ branch. Although it is difficult to analyze fully the complex processes of the underlying electronically resonant process of phonon scattering, here we enumerate some of the key characteristics of these multi-peaked Raman features observed experimentally and provide discussion of the physical origin.

First, the 2ZO'$_N^{(n)}$ peaks generally found to exhibit just two components. This implies that, although the electronic bands in multilayer graphene are numerous, they give rise to electronic processes involving only a small number of dominate phonon momenta in the double-resonance Raman scattering process.

Second, the separation between the two components of the 2ZO'$_N^{(n)}$ peaks decreases with increasing layer number $N$ and decreasing branch index $n$. This trend can be understood as a consequence of the fact that the ZO'$_N^{(n)}$ branches become less dispersive in the relevant region near the zone center with increasing frequency at the $\Gamma$ point. The higher the $\Gamma$-point ZO'-mode frequency, the flatter is the corresponding phonon branch. Since the ZO'-mode $\Gamma$-point frequency increases with increasing $N$ or decreasing $n$, this leads to weaker dispersion of the ZO'$_N^{(n)}$ branch, and, hence, smaller separations between the component peaks of the ZO'$_N^{(n)}$ mode.

Third, the high-frequency components of the LBM tend to be weaker and broader than the low-frequency ones. We may understand this observation by considering the increasingly stronger dispersion of the ZO'-mode phonons at greater distances from the $\Gamma$ point. In the double-resonance Raman process, a larger Raman shift implies a larger wave vectors for the corresponding ZO' phonon. The larger dispersion of such ZO' phonons is then expected to produce a broader feature in the corresponding Raman spectrum.

Fourth, the high-frequency components deviate significantly from the theoretical predictions (Figure 4B-C in the main paper). We take this as an indication of the important role of the selection of the in-plane wave vector in the electronically resonant Raman scattering process.

## 2. 2ZO' Raman band for 3LG with ABA and ABC stacking order

We have directly investigated the influence of the electronic structure on the line shape of the 2ZO' modes in the resonant Raman process. This is accomplished by consideration of both ABA and ABC-stacked trilayer graphene (3LG). For these two types of 3LG samples, the interlayer breathing modes are similar because the lattice coupling between the second nearest neighbor planes is weak. Their electronic structure, however, differs appreciably due to the distinct crystal symmetry (6-12). As shown in Figure S1, the 2ZO' Raman band exhibit double-peak feature for both types of trilayers. The ABC spectrum is, however, found to be broader than the ABA spectrum. In particular, the separation between two components of the ABC spectrum is 10 cm$^{-1}$ larger than that of the ABA spectrum. This indicates that phonons with a larger range of momenta are involved in the ABC system. We attribute this difference to the distinct electronic structure for the ABA and ABC stacked 3LG. This observation is qualitatively consistent with the behavior of the 2D mode (13, 14) and the LOZO' combination mode (15), where the ABC spectra are found to be broader and more structured than the ABA spectra.

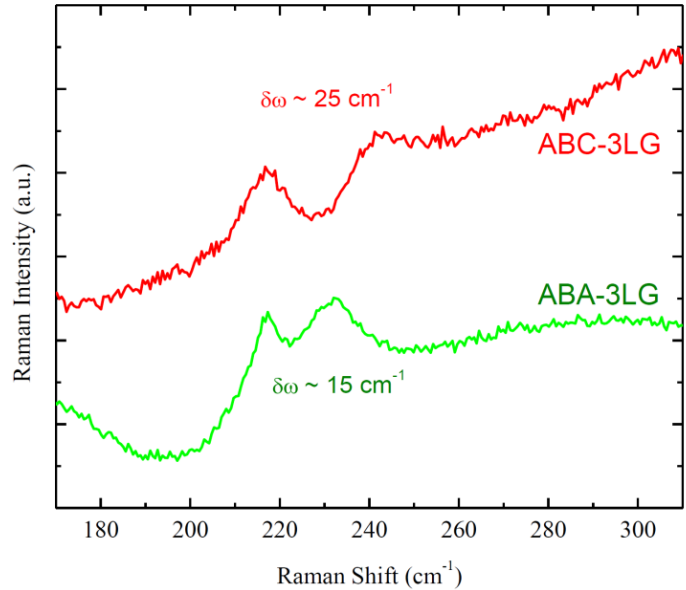

**Figure S1.** The 2ZO'-mode Raman spectra for 3LG with ABA and ABC stacking order. The excitation laser energy (wavelength) is 2.33 eV (532 nm).

**Supplementary References**